\documentclass[a4paper,10pt,twocolumn]{article}
\usepackage[utf8]{inputenc}
\usepackage{multicol,graphicx}
\usepackage{mathptmx}
\usepackage{multirow}
\usepackage{booktabs}
\usepackage{floatpag}
\usepackage{hyperref}
\hypersetup{
    colorlinks=true,
    linkcolor=blue,
    filecolor=magenta,      
    urlcolor=cyan,
    pdfpagemode=FullScreen,
    }

\usepackage[flushleft]{threeparttable}

\usepackage[T1]{fontenc}
\usepackage{textcomp}
\usepackage{url}
\usepackage{dblfloatfix}

\usepackage[T1]{fontenc}
\usepackage[english]{babel}

\usepackage[backend=bibtex,style=ieee,doi=false,isbn=false,url=true,maxnames=6,citestyle=numeric-comp,giveninits=true]{biblatex}

\fontfamily{ptm}\selectfont

\bibliography{references.bib}

\usepackage[font=small]{caption}
\usepackage{amsmath}

\newcommand{\mysection}[1]{\vspace{0.4cm} \uppercase{#1} \vspace{0.4cm}}
\newcommand{\mysubsection}[1]{\hspace{10pt}\textit{#1:}}

\begin{document}
	
\setlength{\textfloatsep}{10pt plus 1.0pt minus 2.0pt}	
\setlength{\columnsep}{1cm}


\twocolumn[%
\begin{@twocolumnfalse}
\begin{center}
	{\fontsize{14}{18}\selectfont
        \textbf{\uppercase{Comparing fingers and gestures for bci control using an optimized classical machine learning decoder}}\\}
    \begin{large}
        \vspace{0.6cm}
        D. Keller\textsuperscript{1}, 
        M. J. Vansteensel\textsuperscript{1},
        S. Mehrkanoon\textsuperscript{2},
        M. P. Branco\textsuperscript{1}\\
        \vspace{0.6cm}
        \textsuperscript{1}UMC Utrecht Brain Center, Department of Neurology and Neurosurgery, Utrecht, The Netherlands\\
        \textsuperscript{2}Department of Information and Computing Sciences, Utrecht University, Utrecht, The Netherlands\\
        \vspace{0.5cm}
        E-mail: \{d.keller, m.pedrosobranco, m.j.vansteensel\}@umcutrecht.nl and s.mehrkanoon@uu.nl
        \vspace{0.4cm}
    \end{large}
\end{center}	
\end{@twocolumnfalse}%
]%


ABSTRACT: Severe impairment of the central motor network can result in loss of motor function, clinically recognized as Locked-in Syndrome. Advances in Brain-Computer Interfaces offer a promising avenue for partially restoring compromised communicative abilities by decoding different types of hand movements from the sensorimotor cortex. In this study, we collected ECoG recordings from 8 epilepsy patients and compared the decodability of individual finger flexion and hand gestures with the resting state, as a proxy for a one-dimensional brain-click. The results show that all individual finger flexion and hand gestures are equally decodable across multiple models and subjects (>98.0\%). In particular, hand movements, involving index finger flexion, emerged as promising candidates for brain-clicks. When decoding among multiple hand movements, finger flexion appears to outperform hand gestures (96.2\% and 92.5\% respectively) and exhibit greater robustness against misclassification errors when all hand movements are included. These findings highlight that optimized classical machine learning models with feature engineering are viable decoder designs for communication-assistive systems.


\mysection{introduction}

Dysfunction of the neuromotor system may precipitate transient or, in severe cases, enduring global loss of motor control. Global dysfunction may be referred to as Locked-In Syndrome (LIS) \cite{plum2000diagnosis}, often characterized by quadriplegia and aphonia. In recent decades, efforts to replace dysfunctional motor control have seen pioneering developments in Brain-Computer Interfaces (BCIs) \cite{thakor2013translating}. BCIs extract information directly from cortical activity to control mechanical or digital effectors without relying on neuromuscular activation, essentially bypassing the muscular output. Restoration of effector control can serve several purposes, ranging from object manipulation \cite{hochberg2012reach, collinger2013high, bouton2016restoring}, locomotion and mobility \cite{courtineBSI2} and speech production \cite{moses2021neuroprosthesis, willett2023high, metzger2023high}. However, for individuals with severe impairment, the restoration of the communicative agency has been identified as one of the most urgent needs \cite{kageyama2020nationwide}.

A simple approach towards communication BCI is automatic letter selection on a digital keyboard \cite{vansteensel2016fully}. An attractive signal recording modality for communication BCIs is electrocorticography (ECoG) due to its high spatiotemporal precision, good signal-to-noise ratio, and reliable signal stability over extended periods \cite{vansteensel2016fully, pels2019stability}. Several studies have demonstrated that hand movement recognition from ECoG recordings can be performed with high accuracy. Consequently, a variety of hand movements have been explored for this purpose, including but not limited to finger flexion \cite{xie2015classifying, xie2018decoding, thomas2020simultaneous, pradeepkumar2021decoding, yao2022fast}, reaching and grasping \cite{hochberg2012reach, collinger2013high, bouton2016restoring, jiang2017characterization, crone2023click}, and wrist flexion and extension \cite{thomas2020simultaneous}, more complex hand gestures \cite{bleichner2016give, branco2017decoding, li2017gesture, thomas2020simultaneous} and handwriting \cite{xu2023swin}. To identify a reliable motor signature for a unidimensional BCI control signal (i.e., 'brain-click') many studies have examined different types of hand movements in isolation  \cite{bleichner2016give, branco2017decoding, li2017gesture, xie2015classifying, xie2018decoding}, but few have compared different hand movements against each other within a unified framework.

\begin{figure}[h]
	\centering
	\includegraphics[width=8cm]{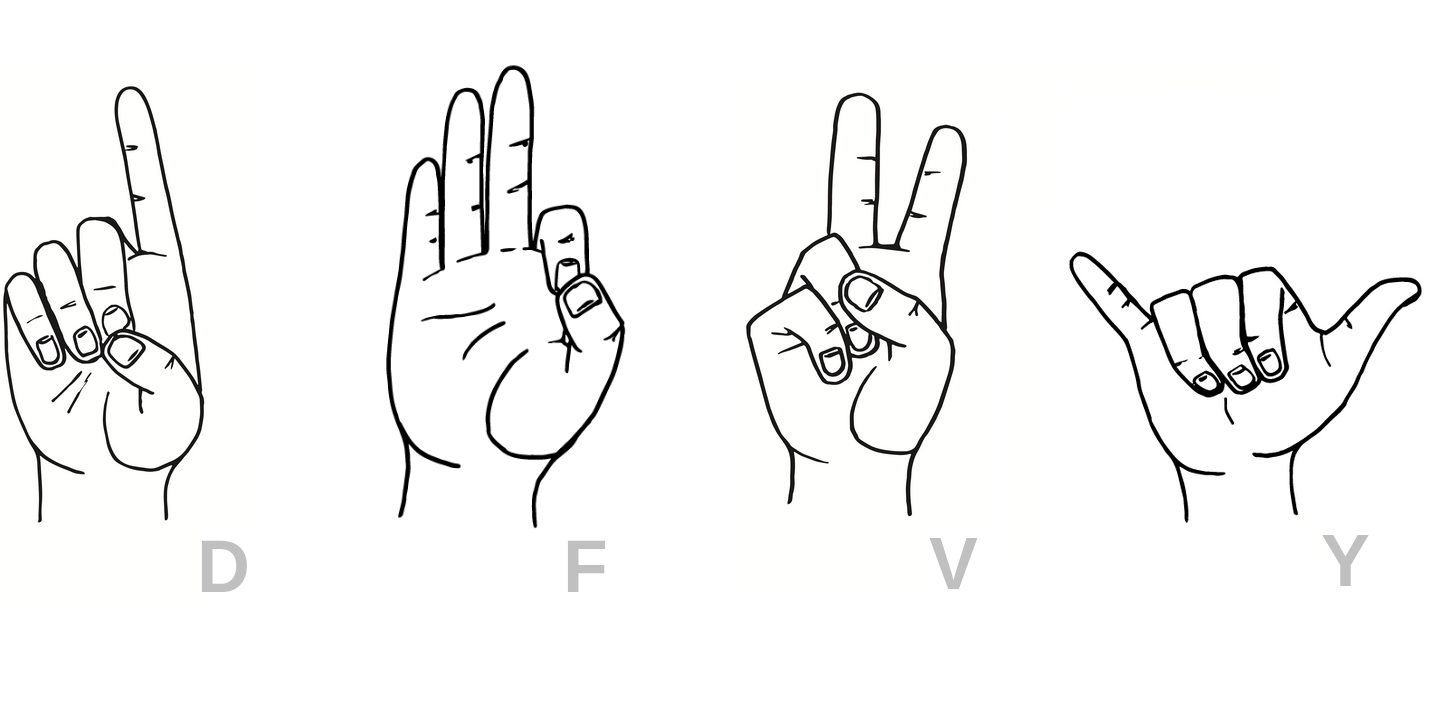}
	\caption{The four gestures executed by subject S1 - S5.}
	\label{fig:gestures}
\end{figure}

The goal of this work is to contribute to a deeper understanding of the decodability of individual finger flexion and hand gestures against the resting state. Specifically, we explored which hand strategy is the most promising for a reliable brain-click and which is more transferable across subjects. In addition, we aim to extend our analysis to the prospect of multidimensional control with four and eight degrees of freedom (DoF) to investigate which type of hand movement intrinsically yields a better within and across-category discriminability. To overcome the notorious data sparsity in this domain, we employed an optimized feature selection decoder with different classification models and assessed which (offline) machine learning approach yields the best performance on individual hand movements (2-DoF), within hand movement types (4-DoF), and within all hand movements (8-DoF).


\mysection{materials and methods}

\mysubsection{Data description} This study is based on two ECoG data sets, consisting of finger flexion, including the index finger, little finger, and thumb, and hand gestures associated with the American Sign Language letters D, F, V, and Y \cite{siero2014bold, branco2017decoding} (see Figure \ref{fig:gestures}).

The data were collected from 8 subjects (\textit{N\textsubscript{G}} = 5, \textit{N\textsubscript{F}} = 6) in an epilepsy monitoring unit of the University Medical Center Utrecht (see Table \ref{tab:subjects}). These subjects had 32, 64 or 128 high-density subdural ECoG electrodes with an inter-electrode distance of 3 or 4mm and an exposed diameter of 1 to 1.3mm (AdTech, Racine, USA; or PMT Corporation, Chanhassen, MN, USA) implemented over the hand-knob region of the sensorimotor cortex. The ECoG data were recorded using a 128-channel Micromed LTM system (subjects 1 - 5; Treviso, Italy; 22 bits, hardware bandpass filter 0.15–134.4Hz; sampling frequency 512Hz) and a Blackrock system (subjects 6 - 8; Microsystems LLC, Salt Lake City, USA, digital bandpass filter 0.3 - 500Hz; sampling frequency 2000Hz). Data were converted to the BIDS standard format \cite{pernet2019eeg}.

\begin{table}[h]
    \begin{threeparttable}
        \caption{Subject Details}
        \label{tab:subjects}
        \begin{footnotesize}
            \begin{tabular}{l*{8}{@{\hspace{4pt}}c}}
                \toprule [1.5pt]
                \textbf{Sub-} & \multirow{2}{*}{\textbf{Task}} & \textbf{Trials}  & \multirow{2}{*}{\textbf{Age}} & \multirow{2}{*}{\textbf{Sex}} & \multirow{2}{*}{\textbf{Hand}} & \textbf{Hand-} & \textbf{Hemi-} & \textbf{Grid}\\
                \textbf{ject} &  & \textbf{(C / T)} & & & & \textbf{edness} & \textbf{sphere} & \textbf{(incl.)}\\
                \midrule
                \multirow{2}{*}{S1} & G & 37 / 74 & \multirow{2}{*}{19} & \multirow{2}{*}{F} & \multirow{2}{*}{Right} & \multirow{2}{*}{Right} & \multirow{2}{*}{Left} & \multirow{2}{*}{4x8 (32)} \\
                & F & 90 / 181 & & & & & & \\[6pt] 
                S2 & G & 68 / 138 & 45 & F & Left & Left & Right & 8x8 (59) \\[3pt] 
                S3 & G & 34 / 69 & 29 & M & Right & Right & Left & 4x8 (29) \\ [3pt]
                \multirow{2}{*}{S4} & G & 32 / 67 & \multirow{2}{*}{19} & \multirow{2}{*}{M} & \multirow{2}{*}{Right} & \multirow{2}{*}{Right} & \multirow{2}{*}{Left} & \multirow{2}{*}{4x8 (31)}  \\
                & F & 90 / 181 & & & & & & \\[3pt] 
                \multirow{2}{*}{S5} & G & 34 / 69 & \multirow{2}{*}{42} & \multirow{2}{*}{M} & \multirow{2}{*}{Right} & \multirow{2}{*}{Right} & \multirow{2}{*}{Left} & \multirow{2}{*}{4x8 (32)} \\
                & F & 88 / 177& & & & & &  \\[3pt] 
                S6 & F & 89 / 179 & 30 & F & Left & Right & Right & 16x8 (123) \\ [3pt]
                S7 & F & 85 / 171 & 20 & F & Right & Right & Left & 8x8 (64)\\ [3pt]
                S8 & F & 84 / 169 & 36 & F & Right & Right & Left & 16x8 (128) \\
                \bottomrule [1.5pt]
            \end{tabular}
        \end{footnotesize}
        \begin{tablenotes}
            \scriptsize
            \item \textit{Note. } Trials are presented as the ratio of hand movement trials per condition (C) out of all trials (T; including the rest trials). In Grid, (incl.) indicates the number of channels included. Abbreviations: Gesture, G; Finger, F, Male, M; Female, F.
        \end{tablenotes}
    \end{threeparttable}
\end{table}
        
\mysubsection{Experimental Design} Subjects were instructed to initiate movements based on visual cues that were presented in a randomized, event-driven design. For the gestures, the subjects imitated the depicted gesture after stimulus onset and maintained the posture until the end of the trial before returning to a resting position. Each subject performed 10 trials with an intertrial interval of 4.4s and a run duration of 6.7m. Rest trials were implicitly calculated from a small time interval before the onset of the next movement. For finger flexion, the subjects performed two finger flexions immediately after cue onset and then returned to a resting position afterward. In contrast to the gestures, each movement was interleaved with an explicit resting trial. The design consisted of 30 trials with an intertrial interval of 7s and a run duration of 8.2m. In both experiments, each subject performed the tasks with the hand contralateral to the grid location, and subject 2 performed the task twice. In addition, a data glove (5DT, Irvine CA, USA, 20 ms sampling time) was used during both experiments to record motor activity.

\mysubsection{Preprocessing} Data preprocessing included the removal of bad trials and channels (identified by \cite{siero2014bold, branco2017decoding} based on data glove data and raw signal inspection), followed by common average referencing, notch, and bandpass filtering (56 Hz - 130 Hz) to remove artifacts. Finger flexion data, sampled at 2000 Hz, were downsampled to 512 Hz for consistency across subjects. The data were then subsequently aligned with movement onset markers obtained from data glove recordings and segmented accordingly. For decoding individual hand movements and within hand movement types a segmentation window of \textit{W\textsubscript{F}} = [-0.5, 1.5s] and \textit{W\textsubscript{G}} = [-0.5s, 2.5s] was used. For decoding all hand movements, the two 4-DoF settings for subject 5 were combined with a common segmentation window of \textit{W\textsubscript{FG}} = [-0.5s, 2s]. In all three settings \textit{t} = 0 represents the motion-aligned stimulus presentation.

Features were extracted using a continuous Morlet wavelet transformation, which produced spectral power features for the high-frequency band (60 Hz - 126 Hz) in 2 Hz frequency bins. To reduce the feature space the power was averaged and the time dimension was decimated to \textit{T\textsubscript{G}} = 154 and \textit{T\textsubscript{F}} = 102 time points per channel for the fingers and gestures, respectively. The resulting feature vectors were used for subsequent model training. Preprocessing was conducted in Python (v3.9) using the MNE library (v1.16).

\begin{figure*}[h]
	\centering
	\includegraphics[width=\linewidth]{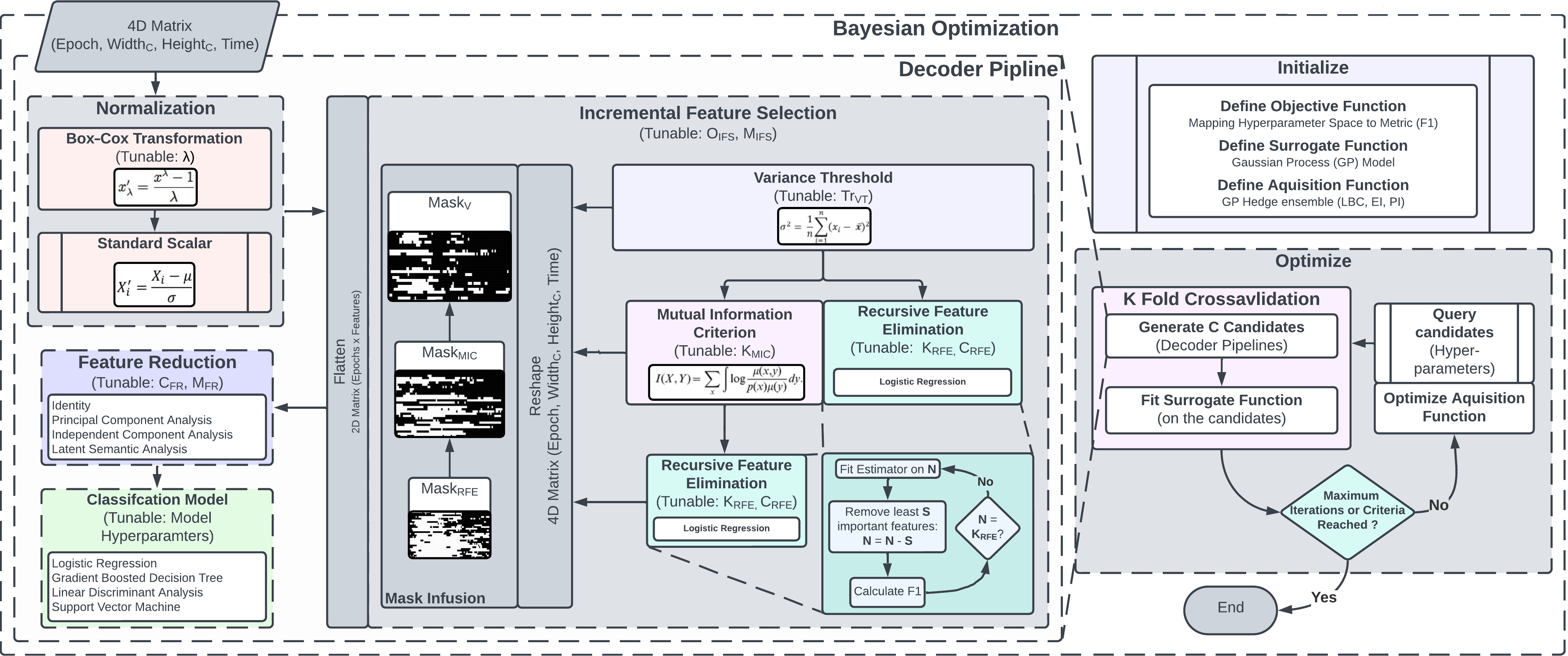}
            \caption{General decoder architecture. The pipeline is composed of four modules: Normalization, Incremental Feature Selection (IFS), Feature Reduction (FR), and a Classification Model. A Box-Cox (BC) transformation and a Standard Scalar normalize the signal. The incremental selection procedure can recruit different combinations of Variance Thresholding (VT), Mutual Information Criterion (MIC), and Recursive Feature Elimination (RFE) to select the relevant features from the spatio-temporal representation. RFE iteratively removes the set of S least important features from the pool N until it reaches K\textsubscript{RFE} features. After feature selection, the feature space can be further compressed with various Feature Reduction (FR) methods. The resulting vector is forwarded to one of four classifiers. A Bayesian optimization algorithm orchestrates the order and method of feature selection (O\textsubscript{IFS}, M\textsubscript{IFS}), other pipeline configuration ($\lambda$\textsubscript{BC},  Tr\textsubscript{VT}, K\textsubscript{MIC}, K\textsubscript{RFE}, C\textsubscript{RFE}, C\textsubscript{FR}), and various model-specific hyperparameters. Dotted lines represent meta-routing processes; single and double-lined boxes represent dynamic and predefined processes, respectively. Abbreviations: LBC, Lower Confidence Bound; EL, Negative Expected Improvement; PI Negative Probability Improvement}
	\label{fig:decoder}
\end{figure*}

\mysubsection{Decoder} The architecture of the decoder, depicted in Figure \ref{fig:decoder}, revolves around an optimized data-driven feature engineering approach for conventional classical machine learning classifiers. The decoder encompasses four modules: Normalization, Incremental Feature Selection (IFS), Feature Reduction (FR), and Classification.

The initial step of the decoder normalises the spectral power of the spatio-temporal feature vector (N) via a Box-Cox transformation \cite{box1964analysis}, followed by mean centering and unit variance scaling to ensure data normality and variance stabilisation. Feature selection employs Variance Thresholding (VT), Mutual Information Criterion (MIC) \cite{kraskov2004estimating, ross2014mutual}, and Recursive Feature Elimination (RFE). RFE iteratively removes a set of features (S = 1e-3 * N) corresponding to the least important coefficients, and akin to MIC, retains a subset of the best K temporal features across all channels. Each selection method generates a binary mask indicating the retained features. The individual modules are applied incrementally, with the specific combination and its parameter configuration being delegated to a Bayesian optimization algorithm, which avoids manual tuning and efficiently navigates through the high-dimensional parameter space. In particular, this approach aims to balance the advantages of filter and wrapper methods \cite{tsamardinos2003towards} to remove noisy and redundant features while prioritising discriminative ones. The Feature Reduction (FR) method can be applied in isolation or in conjunction with IFS processing. In the classification phase, four algorithms were selected to compete with each other: Logistic Regression (LR), Linear Discriminant Analysis (LDA), Boosted Decision Trees (BDT), and Support Vector Machines (SVM). A Majority Class Predictor, which predicts the most frequent hand movement, was evaluated on the data to establish a 'chance' baseline. Performance is assessed using the F1 scoring metric, adjusted for label imbalance with the inversely weighted class distribution. 

For this 'black-box' optimization problem, the Bayesian algorithm \cite{mockus1989bayesian} approximates an expensive non-smooth objective function by inference, essentially guiding the search process based on prior results. To find an optimal decoder candidate in the large parameter space, a Gaussian Process model \cite{rasmussen2006gaussian} with a hedging portfolio strategy \cite{hoffman2011portfolio}is used, where hedging probabilistically choose the best acquisition function from three candidates: Lower Confidence Bound, Negative Expected Improvement or Negative Probability Improvement. The search is restricted to a maximum of 256 candidates, exploring a hyperparameter space, ranging from 9 (LDA) to 20 (BDT) configurations for different algorithms, of which up to 60\% are conditional hyperparameters; the number-of-components hyperperameter for the FR step was shared among all three methods. Model performance is evaluated using stratified 10-fold crossvalidation, with the best candidate further assessed through leave-one-out crossvalidation. The decoder pipeline adheres to the scikit-learn architecture, ensuring compatibility with the scikit-learn library and its derivatives. The implementation is in Python 3.9, using scikit-learn (v1.4.0) and xgboost (v2.0.3).

\mysubsection{Statistical Analysis} The analysis relies on a Friedman ANOVA to identify a general effect and Dunn's test with Benjamini-Hochberg's false discoveries rate correction the post hoc analysis and pairwise comparison. The statistical analysis was performed in Python 3.9, using scipy (v1.13.0) and scikit-posthocs (v0.9.0). 


\mysection{results}

\begin{figure*}[h]
	\centering
	\includegraphics[width=\linewidth]{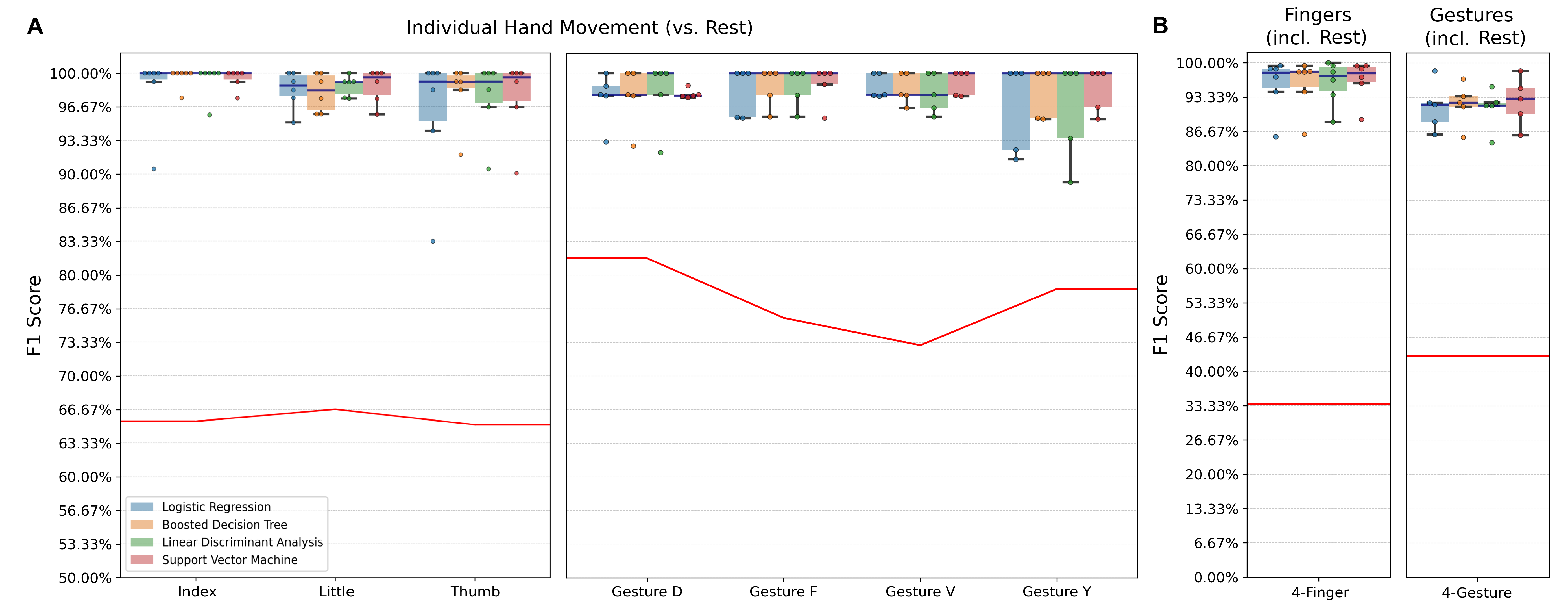}
	\caption{Box plots for different machine learning models of the decoder for (A) all individual fingers and (B) the three multi-DoF comparisons. Each point represents one subject, with the upper and lower error bars representing an interquartile range of 25 and 75, respectively, and where omitted when performance for one or more subjects exceeded this range. The 4-DoF gesture decoding includes Gesture F, Y, and V. The red line represents the highest chance level among all subjects.}
	\label{fig:box_plots}
\end{figure*}

\mysubsection{Individual Hand Movements} In the context of individual hand movements, all classification models exhibited a high F1 performance (averaged across subjects), exceeding 98.0\% for each finger flexion and gesture (details summarised in Table \ref{tab:model_performance} and Figure \ref{fig:box_plots} A). Notably, the Index finger (99.59\%), Gesture V (99.11\%), and Gesture F (98.09\%) were the most promising candidates for brain-click BCI control. Interestingly, they share a commonality in index finger flexion. Within the hand gestures, no gesture significantly outperformed the others (Friedman ANOVA and Dunn's test; ns). Similarly, within the finger flexions, the overall effect was significant, $\chi^2$(3) = \textit{15.32}, p = $4.71e^{-4}$), but no finger flexion was significantly different from the others (Dunn's test). Moreover, decoding performance remained remarkably stable for each subject across all hand movements and models, with consistent trends in variability observed for each subject (i.e., subject 1 consistently had the lowest and highest scores for the fingers and the gestures, respectively). 

\begin{table*}[h]
    \begin{threeparttable}
        \begin{footnotesize}
            \centering
            \caption{Mean performance across all subjects of the Optimize Feature Selection Decoder for Individual Movements (2-DoF), Within Types (4 DoF) and Within all Hand Movements (8-DoF).}
            \label{tab:model_performance}
            \begin{tabular}{l*{20}{@{\hspace{5pt}}c}}
                \toprule [1.5pt]
                \multirow{2}{*}{\textbf{Models}} & \multicolumn{6}{c}{\textbf{Fingers (vs. rest)}} & \multicolumn{8}{c}{\textbf{Gestures (vs. rest)}} & \multicolumn{6}{c}{\textbf{Multi-DoF (incl. rest)}}\\ [1.5pt]
                \cmidrule(lr){2-7} \cmidrule(lr){8-15} \cmidrule(lr){16-21}
                & \multicolumn{2}{c}{\textbf{Index}} & \multicolumn{2}{c}{\textbf{Little}} & \multicolumn{2}{c}{\textbf{Thumb}} & \multicolumn{2}{c}{\textbf{Gesture D}} & \multicolumn{2}{c}{\textbf{Gesture F}} & \multicolumn{2}{c}{\textbf{Gesture V}} & \multicolumn{2}{c}{\textbf{Gesture Y}} & \multicolumn{2}{c}{\textbf{4-Finger}} & \multicolumn{2}{c}{\textbf{4-Gesture\textsuperscript{*}}} & \multicolumn{2}{c}{\textbf{8-Hand\textsuperscript{**}}}\\ [1.5pt]
                \cmidrule(lr){2-3} \cmidrule(lr){4-5} \cmidrule(lr){6-7} \cmidrule(lr){8-9} \cmidrule(lr){10-11} \cmidrule(lr){12-13} \cmidrule(lr){14-15} \cmidrule(lr){16-17} \cmidrule(lr){18-19} \cmidrule(lr){20-21}
                               & A     & F1    & A     & F1    & A     & F1    & A     & F1    & A     & F1    & A     & F1    & A     & F1    & A    & F1   & A    & F1   & A    & F1 \\ [1.5pt]
                \midrule
                Chance         & 75.3 & 64.6 & 75.5 & 64.9 & 74.9 & 64.2 & 83.3 & 75.7 & 79.8 & 70.8 & 79.4 & 70.3 & 79.8 & 70.9 & 50.3 & 33.7 & 56.6 & 40.9 & 50.4 & 33.8 \\ [1.5pt]
                \midrule
                LR             & 98.2 & 98.3 & 98.3 & 98.4 & 95.9 & 96.0 & 97.4 & 97.5 & 98.2 & 98.2 & 98.7 & 98.7 & 96.6 & 96.8 & 95.7 & 95.7 & 91.3 & 91.5 & 87.4 & 87.7 \\ [1.5pt]
                BDT            & 99.6 & \textbf{99.6} & 98.1 & 98.1 & 98.1 & \textbf{98.0} & 97.6 & 97.7 & 98.7 & 98.7 & 98.4  & 98.5 & 98.2 & 98.2 & 95.8 & 95.8 & 92.0 & 92.0 & 88.6 & 88.7 \\ [1.5pt]
                LDA            & 99.3 & 99.3 & 98.7 & \textbf{98.8} & 97.5 & 97.6 & 98.1 & \textbf{98.1} & 98.7 & 98.7 & 98.0 & 98.0 & 96.4 & 96.6 & 96.1 & 96.1 & 91.1 & 91.1 & 93.1 & \textbf{92.9} \\ [1.5pt]
                SVM            & 99.5 & 99.4 & 98.8 & \textbf{98.8} & 97.7 & 97.6 & 97.9 & 98.0 & 98.8 & \textbf{98.89} & 99.1 & \textbf{99.1} & 98.4 & \textbf{98.4} & 96.7 & \textbf{96.6} & 92.5 & \textbf{92.5} & 92.7 & 92.6 \\ [1.5pt]
                \bottomrule [1.5pt]
            \end{tabular}
        \end{footnotesize}
        \begin{tablenotes}
            \scriptsize
            \item \textit{Note. } Values are in \%. \textsuperscript{*}The 4-DoF gesture decoding includes Gesture F, Y, and V. \textsuperscript{**} The 8-DoF decoding of all hand movements was only obtained from subject 5. Abbreviations: Accuracy, A; Versus, vs; Inclusive, incl.
        \end{tablenotes}
    \end{threeparttable}
\end{table*}

\begin{figure*}[h]
	\centering
	\includegraphics[width=\linewidth]{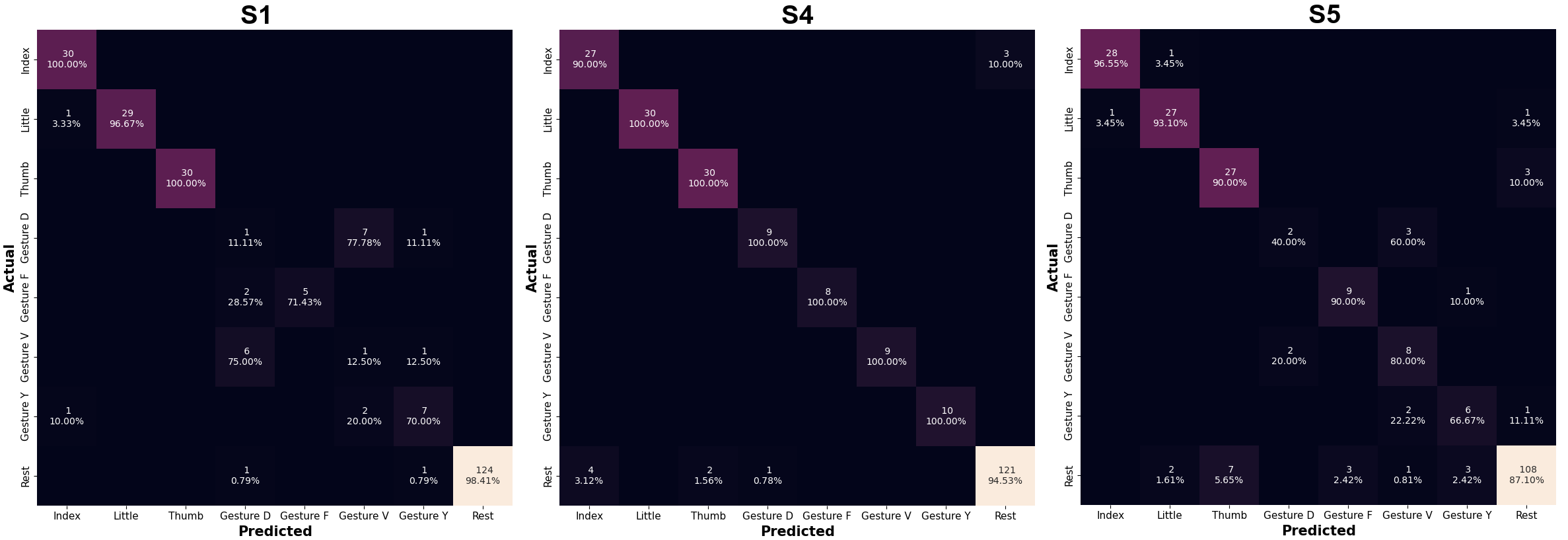}
	\caption{Confusion matrices of the LR model for subjects 1, 4, and 5. In each confusion matrix, the horizontal axis represents the predicted hand movement (or rest), and the vertical axis represents the ground truth hand movement (or rest). Henceforth, the diagonal elements represent the total correct values predicted per hand movement. The colour coding represents the proportion of absolute counts for each hand movement ranging from black (0\%) to white (100\%).}
	\label{fig:confusion_matrices}
\end{figure*}

\mysubsection{Hand Movements Types} For the hand movement types, a different trend emerged. For 4-DoF classification, finger flexion (95.7\% - 96.6\%) outperformed the hand gestures (91.1\% - 92.5\%; average across subjects), which could be statistically verified (Dunn's test), p = 0.031; we excluded the worst decodable gesture (Gesture D) to ensure a similar task complexity. An extension to 8-DoF classification preserves a high F1 score for Subject 5 (88.7\% - 92.9\%), with remarkably minimal confusion between gestures and fingers. On visual inspection, fingers exhibit more confusion with the resting state, while gestures are more often confused among themselves (as depicted in Figure \ref{fig:confusion_matrices}).

\mysubsection{Classification Models} When training classification models within an optimised feature engineering framework, no model emerges as significantly superior to the others (Friedman ANOVA), $\chi^2$(4) = \textit{60.48}, p = {$2.3e^{-12}$}, (Dunn's test; ns), although, all perform significantly above chance level (Dunn's Test), p < $1.28e^{-8}$. In general, Boosted Decision Trees (BDT) and Support Vector Machines (SVM) demonstrate the highest classification performance across all conditions, except for Little Finger and Gesture D. 


\mysection{discussion}

The current work demonstrates that optimised spatio-temporal feature engineering of finger flexion and hand gestures, recorded from high-density ECoG, enables reliable decoding for one-dimensional brain-click, 4- and 8-DoF decoding tasks, even with very small data volumes. Notably, within each category, no single hand movement emerged as superior decodable. However, upon qualitative inspection, the index finger and Gestures V and F, all sharing index finger flexion, appeared as the most promising candidates. Moreover, the fingers exhibited a better performance in more complex 4- and 8-DoF decoding; finger flexion may possess more discriminative properties for multi-DoF tasks. Finally, no classical machine learning model outperformed the others, but BDT and SVM may have a small advantage. 

Hand gestures entail a more complex interplay of motor components than simple finger flexion, including wrist and finger flexion, and lateral extension, among others. However, our results revealed that individual finger flexion alone yields a near perfect neuroelectrical signature. Consequently, the addition of supplementary motor components may not increase decodability but rather confront the decoder with a motor signature that inherently has a higher variability in its signal. As we evaluated the decoder on small data volumes, an effect of additional components on decoder performance may emerge with larger sample sizes. Extending this rationale, gestures might possess a more intricate spatial and temporal pattern, recruiting various neuronal populations from a larger effector field in the sensorimotor cortex. The absence of advanced feature transformation techniques in our classical machine learning approach might hinder the decoder from exploiting the full potential of the gesture's electrical signature. In addition, differences in sample size, with fewer than 30 trials for fingers and only 10 trials for gestures, may impact the comparison, while the overall small sample size may not provide sufficient power to detect a potential statistical difference.

In line with the evident discernible difference between the 4-DoF types, misclassification errors were more pronounced for gestures, with frequent confusion among different fingers and the resting state. Notably, substantial confusion between hand movements and the resting state suggests a potential contamination of the rest periods with unintended movement. This may be attributed to the design of the experimental design of the gestures, which lacked separate explicit rest trials. Incorporating a threshold based on data glove recordings could be beneficial. However, defining true rest may not be practically feasible for real BCI applications, as it necessitates extensive subject training to suppress such activity \cite{vansteensel2016fully, pels2019stability, crone2023click}, and a more naturalistic approach would be to build a decoder that can successfully discriminate between meaningful and non-task-related sporadic motor activity. Furthermore, some features along the temporal dimension may not reflect actual motor activity, but 'resting' activity, especially before the movement onset and towards the end of the segmentation window.

The proposed decoder design surpassed prior approaches evaluated on gesture data for four out of five subjects \cite{bleichner2016give, branco2017decoding}, albeit a direct comparison is difficult due to differences in task complexity. For communication assistance systems, the proposed decoder design can offer a viable alternative to deep learning approaches for one-dimensional brain-click tasks \cite{crone2023click} and even larger DoF applications \cite{xie2018decoding, yao2022fast}, where data acquisition is challenging. Although trained offline, the decoder can process individual segments of preprocessed data in as little as 2 - 10ms. Consequently, although theoretically deployable in an online setting, regular offline retraining on new data is necessary to address concept drifts for ensuring long-term stability, in particular for individuals with neurodegenerative diseases. Importantly, the experiments involved movement execution by epileptic individuals, further validation in attempted movement is imperative to extend applicability to online BCI settings for individuals with LIS.


\mysection{conclusion}

Electrocorticography data provides a high-resolution spatiotemporal feature representation, forming a suitable foundation to tailor an optimised classical machine learning decoder with automatic feature engineering to the large feature space. We demonstrate that within this framework, both finger flexion and hand gestures enable reliable decoding across multiple subjects, and when extended to a multiple degrees of freedom, maintain high discriminability between hand movements. For click-based letter selection in communication-assistive BCI systems, the index finger flexion emerges as an optimal candidate. Moreover, all tested models consistently exhibit high classification performance across multiple subjects - a comparable performance to deep learning approaches.

\newpage
\mysection{references}
\printbibliography[heading=none]

@book{plum2000diagnosis, 
  title={The diagnosis of stupor and coma},
  author={Plum, Fred and Posner, Jerome B},
  volume={19},
  year={2000},
  publisher={Oxford University Press, USA},
  edition={3rd Edition},
}

@article{thakor2013translating,
  title={Translating the brain-machine interface},
  author={Thakor, Nitish V},
  journal={Science translational medicine},
  volume={5},
  number={210},
  pages={210ps17--210ps17},
  year={2013},
  publisher={American Association for the Advancement of Science}
}

@article{hochberg2012reach,
  title={Reach and grasp by people with tetraplegia using a neurally controlled robotic arm},
  author={Hochberg, Leigh R and Bacher, Daniel and Jarosiewicz, Beata and Masse, Nicolas Y and Simeral, John D and Vogel, Joern and Haddadin, Sami and Liu, Jie and Cash, Sydney S and Van Der Smagt, Patrick and others},
  journal={Nature},
  volume={485},
  number={7398},
  pages={372--375},
  year={2012},
  publisher={Nature Publishing Group UK London}
}

@article{collinger2013high,
  title={High-performance neuroprosthetic control by an individual with tetraplegia},
  author={Collinger, Jennifer L and Wodlinger, Brian and Downey, John E and Wang, Wei and Tyler-Kabara, Elizabeth C and Weber, Douglas J and McMorland, Angus JC and Velliste, Meel and Boninger, Michael L and Schwartz, Andrew B},
  journal={The Lancet},
  volume={381},
  number={9866},
  pages={557--564},
  year={2013},
  publisher={Elsevier}
}

@article{bouton2016restoring,
  title={Restoring cortical control of functional movement in a human with quadriplegia},
  author={Bouton, Chad E and Shaikhouni, Ammar and Annetta, Nicholas V and Bockbrader, Marcia A and Friedenberg, David A and Nielson, Dylan M and Sharma, Gaurav and Sederberg, Per B and Glenn, Bradley C and Mysiw, W Jerry and others},
  journal={Nature},
  volume={533},
  number={7602},
  pages={247--250},
  year={2016},
  publisher={Nature Publishing Group UK London}
}

@article{courtineBSI2,
  title={Walking naturally after spinal cord injury using a brain--spine interface},
  author={Lorach, Henri and Galvez, Andrea and Spagnolo, Valeria and Martel, Felix and Karakas, Serpil and Intering, Nadine and Vat, Molywan and Faivre, Olivier and Harte, Cathal and Komi, Salif and others},
  journal={Nature},
  pages={1--8},
  year={2023},
  publisher={Nature Publishing Group UK London}
}

@article{moses2021neuroprosthesis,
  title={Neuroprosthesis for decoding speech in a paralyzed person with anarthria},
  author={Moses, David A and Metzger, Sean L and Liu, Jessie R and Anumanchipalli, Gopala K and Makin, Joseph G and Sun, Pengfei F and Chartier, Josh and Dougherty, Maximilian E and Liu, Patricia M and Abrams, Gary M and others},
  journal={New England Journal of Medicine},
  volume={385},
  number={3},
  pages={217--227},
  year={2021},
  publisher={Mass Medical Soc}
}

@article{willett2023high,
  title={A high-performance speech neuroprosthesis},
  author={Willett, Francis R and Kunz, Erin M and Fan, Chaofei and Avansino, Donald T and Wilson, Guy H and Choi, Eun Young and Kamdar, Foram and Glasser, Matthew F and Hochberg, Leigh R and Druckmann, Shaul and others},
  journal={Nature},
  volume={620},
  number={7976},
  pages={1031--1036},
  year={2023},
  publisher={Nature Publishing Group UK London}
}

@article{metzger2023high,
  title={A high-performance neuroprosthesis for speech decoding and avatar control},
  author={Metzger, Sean L and Littlejohn, Kaylo T and Silva, Alexander B and Moses, David A and Seaton, Margaret P and Wang, Ran and Dougherty, Maximilian E and Liu, Jessie R and Wu, Peter and Berger, Michael A and others},
  journal={Nature},
  volume={620},
  number={7976},
  pages={1037--1046},
  year={2023},
  publisher={Nature Publishing Group UK London}
}

@article{kageyama2020nationwide,
  title={Nationwide survey of 780 Japanese patients with amyotrophic lateral sclerosis: their status and expectations from brain--machine interfaces},
  author={Kageyama, Yu and He, Xin and Shimokawa, Toshio and Sawada, Jinichi and Yanagisawa, Takufumi and Shayne, Morris and Sakura, Osamu and Kishima, Haruhiko and Mochizuki, Hideki and Yoshimine, Toshiki and others},
  journal={Journal of Neurology},
  volume={267},
  pages={2932--2940},
  year={2020},
  publisher={Springer}
}

@article{vansteensel2016fully,
  title={Fully implanted brain--computer interface in a locked-in patient with ALS},
  author={Vansteensel, Mariska J and Pels, Elmar GM and Bleichner, Martin G and Branco, Mariana P and Denison, Timothy and Freudenburg, Zachary V and Gosselaar, Peter and Leinders, Sacha and Ottens, Thomas H and Van Den Boom, Max A and others},
  journal={New England Journal of Medicine},
  volume={375},
  number={21},
  pages={2060--2066},
  year={2016},
  publisher={Mass Medical Soc}
}

@article{pels2019stability,
  title={Stability of a chronic implanted brain-computer interface in late-stage amyotrophic lateral sclerosis},
  author={Pels, Elmar GM and Aarnoutse, Erik J and Leinders, Sacha and Freudenburg, Zac V and Branco, Mariana P and van der Vijgh, Benny H and Snijders, Tom J and Denison, Timothy and Vansteensel, Mariska J and Ramsey, Nick F},
  journal={Clinical Neurophysiology},
  volume={130},
  number={10},
  pages={1798--1803},
  year={2019},
  publisher={Elsevier}
}

@article{xie2015classifying,
  title={Classifying multiple types of hand motions using electrocorticography during intraoperative awake craniotomy and seizure monitoring processes—case studies},
  author={Xie, Tao and Zhang, Dingguo and Wu, Zehan and Chen, Liang and Zhu, Xiangyang},
  journal={Frontiers in neuroscience},
  volume={9},
  pages={353},
  year={2015},
  publisher={Frontiers Media SA}
}

@article{xie2018decoding,
  title={Decoding of finger trajectory from ECoG using deep learning},
  author={Xie, Ziqian and Schwartz, Odelia and Prasad, Abhishek},
  journal={Journal of neural engineering},
  volume={15},
  number={3},
  pages={036009},
  year={2018},
  publisher={IOP Publishing}
}

@article{thomas2020simultaneous,
  title={Simultaneous classification of bilateral hand gestures using bilateral microelectrode recordings in a tetraplegic patient},
  author={Thomas, Tessy M and Nickl, Robert W and Thompson, Margaret C and Candrea, Daniel N and Fifer, Matthew S and McMullen, David P and Osborn, Luke E and Pohlmeyer, Eric A and Anaya, Manuel and Anderson, William S and others},
  journal={MedRxiv},
  pages={2020--06},
  year={2020},
  publisher={Cold Spring Harbor Laboratory Press}
}

@inproceedings{pradeepkumar2021decoding,
  author={Pradeepkumar, Jathurshan and Anandakumar, Mithunjha and Kugathasan, Vinith and Lalitharatne, Thilina D. and De Silva, Anjula C. and Kappel, Simon L.},
  booktitle={2021 43rd Annual International Conference of the IEEE Engineering in Medicine and Biology Society (EMBC)}, 
  title={Decoding of Hand Gestures from Electrocorticography with LSTM Based Deep Neural Network}, 
  year={2021},
  pages={420-423},
}

@article{yao2022fast,
  title={Fast and accurate decoding of finger movements from ECoG through Riemannian features and modern machine learning techniques},
  author={Yao, Lin and Zhu, Bingzhao and Shoaran, Mahsa},
  journal={Journal of Neural Engineering},
  volume={19},
  number={1},
  pages={016037},
  year={2022},
  publisher={IOP Publishing}
}

@article{jiang2017characterization,
  title={Characterization and decoding the spatial patterns of hand extension/flexion using high-density ECoG},
  author={Jiang, Tianxiao and Jiang, Tao and Wang, Taylor and Mei, Shanshan and Liu, Qingzhu and Li, Yunlin and Wang, Xiaofei and Prabhu, Sujit and Sha, Zhiyi and Ince, Nuri F},
  journal={IEEE Transactions on Neural Systems and Rehabilitation Engineering},
  volume={25},
  number={4},
  pages={370--379},
  year={2017},
  publisher={IEEE}
}

@article{crone2023click,
  title={A click-based electrocorticographic brain-computer interface enables long-term high-performance switch-scan spelling},
  author={Crone, Nathan and Candrea, Daniel and Shah, Samyak and Luo, Shiyu and Angrick, Miguel and Rabbani, Qinwan and Coogan, Christopher and Milsap, Griffn and Nathan, Kevin and Wester, Brock and others},
  journal={Research Square},
  year={2023},
  publisher={American Journal Experts}
}

@article{bleichner2016give,
  title={Give me a sign: decoding four complex hand gestures based on high-density ECoG},
  author={Bleichner, Martin G and Freudenburg, Zachary V and Jansma, Johannus Martijn and Aarnoutse, Erik J and Vansteensel, Mariska J and Ramsey, Nick Franciscus},
  journal={Brain Structure and Function},
  volume={221},
  pages={203--216},
  year={2016},
  publisher={Springer}
}

@article{branco2017decoding,
  title={Decoding hand gestures from primary somatosensory cortex using high-density ECoG},
  author={Branco, Mariana P and Freudenburg, Zachary V and Aarnoutse, Erik J and Bleichner, Martin G and Vansteensel, Mariska J and Ramsey, Nick F},
  journal={NeuroImage},
  volume={147},
  pages={130--142},
  year={2017},
  publisher={Elsevier}
}

@article{li2017gesture,
  title={Gesture decoding using ECoG signals from human sensorimotor cortex: a pilot study},
  author={Li, Yue and Zhang, Shaomin and Jin, Yile and Cai, Bangyu and Controzzi, Marco and Zhu, Junming and Zhang, Jianmin and Zheng, Xiaoxiang and others},
  journal={Behavioural neurology},
  volume={2017},
  year={2017},
  publisher={Hindawi}
}

@article{xu2023swin,
  title={Swin-TCNet: Swin-based temporal-channel cascade network for motor imagery iEEG signal recognition},
  author={Xu, Mingyue and Zhou, Wenhui and Shen, Xingfa and Wang, Yuhan and Mo, Liangyan and Qiu, Junping},
  journal={Biomedical Signal Processing and Control},
  volume={85},
  pages={104885},
  year={2023},
  publisher={Elsevier}
}

@article{siero2014bold,
  title={BOLD matches neuronal activity at the mm scale: a combined 7 T fMRI and ECoG study in human sensorimotor cortex},
  author={Siero, Jeroen CW and Hermes, Dora and Hoogduin, Hans and Luijten, Peter R and Ramsey, Nick F and Petridou, Natalia},
  journal={Neuroimage},
  volume={101},
  pages={177--184},
  year={2014},
  publisher={Elsevier}
}

@article{pernet2019eeg,
  title={EEG-BIDS, an extension to the brain imaging data structure for electroencephalography},
  author={Pernet, Cyril R and Appelhoff, Stefan and Gorgolewski, Krzysztof J and Flandin, Guillaume and Phillips, Christophe and Delorme, Arnaud and Oostenveld, Robert},
  journal={Scientific data},
  volume={6},
  number={1},
  pages={103},
  year={2019},
  publisher={Nature Publishing Group UK London}
}

@article{box1964analysis,
  title={An analysis of transformations},
  author={Box, George EP and Cox, David R},
  journal={Journal of the Royal Statistical Society Series B: Statistical Methodology},
  volume={26},
  number={2},
  pages={211--243},
  year={1964},
  publisher={Oxford University Press}
}

@article{kraskov2004estimating,
  title={Estimating mutual information},
  author={Kraskov, Alexander and St{\"o}gbauer, Harald and Grassberger, Peter},
  journal={Physical review E},
  volume={69},
  number={6},
  pages={066138},
  year={2004},
  publisher={APS}
}

@inproceedings{tsamardinos2003towards,
  title={Towards principled feature selection: Relevancy, filters and wrappers},
  author={Tsamardinos, Ioannis and Aliferis, Constantin F},
  booktitle={International Workshop on Artificial Intelligence and Statistics},
  pages={300--307},
  year={2003},
  organization={PMLR}
}

@article{ross2014mutual,
  title={Mutual information between discrete and continuous data sets},
  author={Ross, Brian C},
  journal={PloS one},
  volume={9},
  number={2},
  pages={e87357},
  year={2014},
  publisher={Public Library of Science San Francisco, USA}
}

@book{rasmussen2006gaussian,
  title={Gaussian processes for machine learning},
  author={Rasmussen, Carl Edward and Williams, Christopher KI and others},
  volume={1},
  year={2006},
  publisher={Springer}
}

@book{mockus1989bayesian,
  title={The Bayesian approach to local optimization},
  author={Mockus, Jonas and Mockus, Jonas},
  year={1989},
  publisher={Springer}
}

@inproceedings{hoffman2011portfolio,
  title={Portfolio Allocation for Bayesian Optimization.},
  author={Hoffman, Matthew and Brochu, Eric and De Freitas, Nando and others},
  booktitle={UAI},
  pages={327--336},
  year={2011}
}
\end{document}